\documentclass[aps,twocolumn,prb]{revtex4-1}
\usepackage{longtable}
\usepackage{subfigure}
\usepackage{amsmath,amssymb,amsfonts,bm}
\usepackage{graphicx}
\usepackage{ulem}
\usepackage{color}

\def\KeyWord#1{$\backslash$\IfColor{$\!\!$\textRed{#1}\textBlack}{#1}$\!\!$}
\newcommand{\be}{\begin{equation} }
\newcommand{\ee}{\end{equation} }
\newcommand{\ba}{\begin{eqnarray} }
\newcommand{\ea}{\end{eqnarray} }
\newcommand{\n}{\nonumber \\ }

\newcommand{\bit}{\begin{itemize}}
\newcommand{\eit}{\end{itemize}}

\begin{document}
\title{Correlated parity measurements as a probe of non-abelian statistics in 1D superconducting wires}

\author{F. J.  Burnell}
\affiliation{All Souls College, Oxford, OX14AL, UK}
\affiliation{Rudolf Peirels center for theoretical physics, Oxford, UK}

\begin{abstract}
We propose a method to detect a signature of non-abelian statistics in a 1D superconducting wire by tuning the effective coupling between a pair of Majorana particles.  
Our experiment requires a single wire with two segments in the topological superconducting phase, and a total of 4 Majorana particles.  We show that an appropriately timed ``pulse" in the coupling between the two middle Majoranas leads to a coherent rotation of the two q-bits associated with the pairs of Majoranas in the two TSC segments, in much the same way that an appropriate-length pulse of a transverse magnetic field can be used to rotate spins in NMR.  This can be exploited both to probe the correlations of the Majoranas, and to manipulate the Majorana q-bits, in these 1D wires. 
We discuss the experimental requirements for such a coherent rotation to be observable when the wire system is coupled to a fermionic bath.  
\end{abstract}

\maketitle

\section{Introduction}
Realizing non-Abelian statistics in condensed matter systems is an exciting prospect, both from the point of view of fundamental theory, and for possible applications to quantum computing\cite{NayakReview}.  Though the possibility of non-abelian statistics is well-established in the context of quantum Hall physics, only recently\cite{KitaevMajorana,OregPRL105.177002,LutchynPRL105.077001} was it realized that 1D superconducting wires with large spin-orbit coupling also have the potential to host particles with non-abelian statistics\cite{MooreRead} known as Majorana zero modes.  
Experimental progress towards realizing these systems has been extremely promising: several groups\cite{MourikScience336,DengNanoLett12,DasNatPhys8}
have observed tunneling conductance peaks at zero bias  in appropriately engineered nanowires, whose dependence on the applied magnetic field closely matches theoretical predictions.  

Observing such zero-bias tunneling peaks at the ends of these 1D topological superconducting (TSC) wires, however, leaves room for doubt as to whether or not the zero-energy excitations causing them have the sought-after non-abelian  statistics\cite{RoyTewari}; a direct probe of these statistics would be extremely gratifying for experimentalists and theorists alike.  Unfortunately, a bullet-proof demonstration of such statistics requires carrying out a braiding operation, which requires either a tri-junction geometry\cite{AliceaNatPhys7,SauPRB84},  
or coupling to Josephson junctions or quantum dots\cite{FuKanePRB79,SauPRA82,FlensbergPRL106,vanHeckNJP14}.  Here we propose an alternative experimental protocol, inspired by the principles of measurement-only topological quantum computation\cite{BondersonMeasurement}, which can be realized in a single nanowire.   The gist of our proposal is that  the effects of non-Abelian statistics can be mimicked by varying in time the coupling between two pairs of Majoranas, in an experimental set-up potentially simpler than is required for a  braiding experiment.  We will call our proposed experiment the {\it Majorana Parity Resonance} (MPR) experiment.

The key advance required to carry out our experiment is the ability to measure the fermion parity (which encapsulates the collective state of the two Majorana zero modes at the end-points of a TSC segment) simultaneously in two segments of TSC wire.  In quantum computing parlance, this means that we must be able to read out the $q$-bit on each wire segment.  While no experimental group has yet achieved this, multiple proposals can be found in the literature\cite{BondersonPRL106,HasslerNJP12,HasslerNJP13,Blais2013,YuvalMeasurement}  for how to carry out such a measurement.  Hence there is reason to be optimistic that it will be possible in the near future to perform such a read-out measurement without dramatically increasing the relaxation rates of the Majorana zero modes.

 \begin{figure}
 \includegraphics[width=1.0\linewidth]{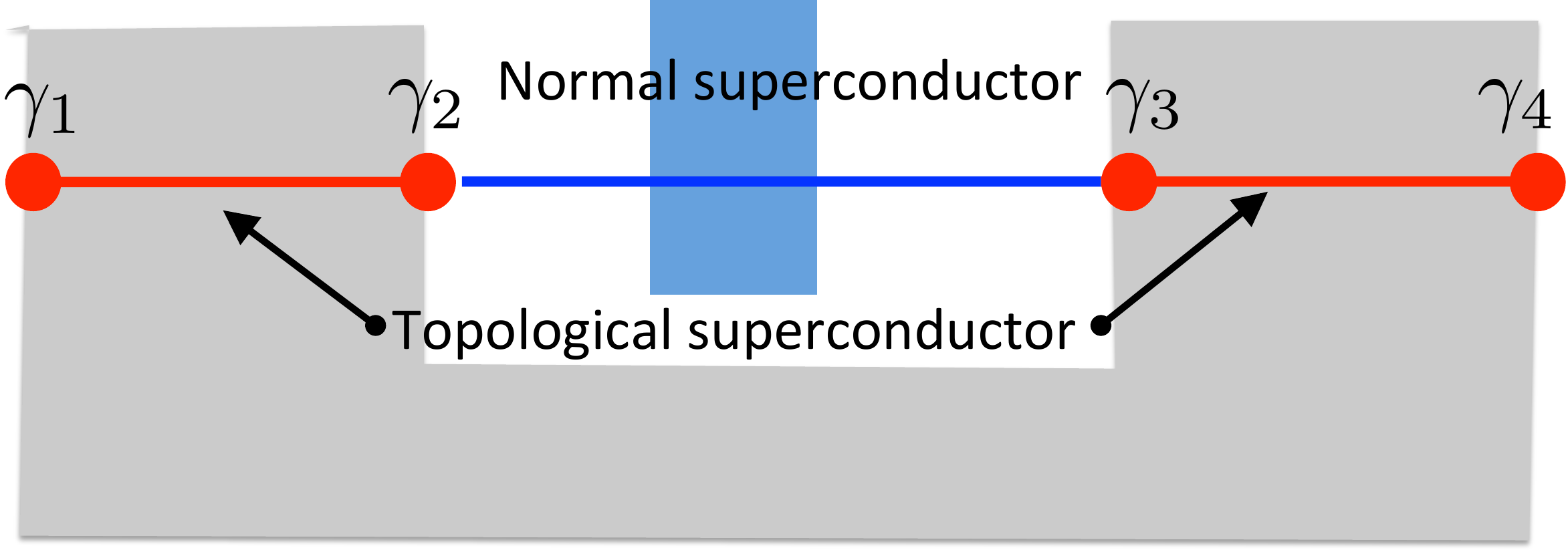} 
  \caption{ Experimental set-up for 1D superconducting wires required for  the MPR experiment.  Red lines indicate segments of the wire that are in the topological superconducting state; blue lies indicate segments in the normal superconducting state.  The red dot at the boundary between normal and topological superconducting regions indicates the candidate Majorana zero mode.   The set-up requires two gates: one to tune the two end segments into the TSC regime (shown here in grey) and one that is used to tune the coupling between $\gamma_2$ and $\gamma_3$ (shown here in blue).}
\label{WireFig}
 \end{figure}

The experimental set-up is shown in Fig. \ref{WireFig}.   
In the MPR experiment, the fermion parity is measured simultaneously in the right-and left-hand wire segments, while the chemical potential in the middle (normal) segment is varied.  
Roughly speaking, the experiment proceeds as follows:  we initialize the system into a state with a definite, known fermion parity in the two wire segments, and with a very large chemical potential in the middle segment, such that the two wire segments are essentially uncoupled.  After some time we apply a brief pulse that lowers the chemical potential in the middle segment, coupling the fermion parities in the two TSC segments.  This perturbation causes the two fermion parities to precess, such that an appropriately timed pulse will bring the system from (say) a state where both fermion parities are $1$ to a state where with a high probability we will measure both to be $0$.  By measuring the two fermion parities simultaneously shortly before and shortly after the pulse, this oscillation can be detected, giving extremely compelling evidence for the entanglement characteristic of non-abelian statistics.  

Our proposal builds on the idea\cite{ParityTime,AguadoPRL} that high-frequency dynamical measurements can harbor signatures characteristic of non-abelian statistics.   Indeed, the same set-up\cite{ParityTime} can be used to demonstrate that the fermion parities in the left- and right- wire segments are correlated.  The MPR experiment, while technically more challenging, goes one step beyond this: we will show that the fact that the fermion parities precess is the dynamical analogue of a process that exchanges two of the Majoranas through a series of measurements.   Because of this connection, the methods described here can also be used to perform high-frequency manipulations of entangled pairs of topological q-bits.  (Conversely, our results also highlight the fact that in the regime we consider here, the 1D wire systems do {\it not} behave topologically).  

One of the advantages of a dynamical measurement over an actual braiding experiment is that the it can be performed on short time-scales, making the signal much more robust to parity relaxation processes. As we will discuss in more detail below, even using a relatively pessimistic estimate of the relaxation rate\cite{RainisLossPRB85.174533}, we find that for the realistic estimates\cite{RainisPRBB87,DasSarmaSmoking}  of the parameters in our model, the minimum pulse width ($\sim 1 \ ns$) is approximately two orders of magnitude faster than the relaxation time ($\sim 100 \ ns$).  This suggests that the signal should be measurable provided that the parity  can be measured quickly relative to the relaxation time -- a requirement that any experimental probe of non-abelian statistics will have to meet.  
It also follows that our approach can be used to coherently rotate pairs of Majorana q-bits on time-scales that can realistically be expected to be fast relative to the predicted relaxation rate.

We note that  variants on the experimental geometry shown in Fig. \ref{WireFig} are possible.  For example, we can replace the middle, normal superconducting segment of the wire by a superconucting quantum dot, which is itself Josephson-coupled to a bulk superconductor.\cite{vanHeckNJP14}  In such a set-up the coupling $t_{23}$ is tuned by the flux through the Josephson junction, rather than by a gate.

The remainder of this work is structured as follows.  In Section \ref{MPRSec}, we will describe the experiment in more detail, explaining the origin of the coherent parity rotation in a simple model of the wire system.  Section \ref{SomeNumerics} presents numerical results for the dynamics for a range of parameter values, and discusses which regime is germane to current experiments.  Section \ref{MeasurementOnlySec} describes the connection between the coherent parity rotation that our experiment probes and the non-abelian statistics of the Majorana zero modes.  We conclude with a brief summary of our findings.

\section{The MPR Experiment} \label{MPRSec}

In this section we will describe the MPR experiment in more detail.  We present an idealized theoretical model of the system, which consists of $4$ Majorana fermions which we couple to a bath to simulate a finite relaxation time.  In the next section we will present numerical results of the dynamics of this system, and discuss the constraints on experimental parameters required in order for a signal to be detected.  

Let us begin with a brief review of the TSC system, to clarify our assumptions and notation.  In practise, the set-up in Fig. \ref{WireFig} consists of a single nanowire in which superconductivity is induced by proximity to a bulk 3D superconductor.  Throughout this work we restrict our attention to energy scales below the induced superconducting gap; in this regime the Hilbert space of the wire consists of the four Majorana zero modes $\gamma_1, ... \gamma_4$ pinned to the endpoints of the two TSC segments.
However, it is important for our purposes that the coherence length of the 3D bulk superconductor is longer than the nanowire; otherwise the required coherence between the pairs of Majorana fermions will not be achieved.

To split our wire into the configuration shown in Fig. \ref{WireFig}, we exploit the fact that the transition between TSC and NTSC wires can be tuned by adjusting either the chemical potential or the magnetic field\cite{OregPRL105.177002,LutchynPRL105.077001}.  (Here we will describe our experiment in terms of the former variable, although either could potentially be used.)  
Our set-up requires two gates: a two-pronged gate with an arm under each of the two end segments of the wire, and a second gate under the central NTSC segment.  The chemical potential of the two-pronged gate is kept fixed at a value that maintains the two end segments in the TSC regime.  The chemical potential of the central gate will be pulsed between a high value at which the two TSC segments are effectively decoupled, and a lower value at which the coupling between the Majoranas $\gamma_2$ and $\gamma_3$ is comparable to the coupling between $\gamma_1$ and $\gamma_2$ (which we take to be approximately equal to the coupling between $\gamma_3$ and $\gamma_4$).  

At energies well below the superconducting gap, the Hamiltonian for this set-up is: 
\be  \label{HgEq}
H(t) = t_{12} \gamma_1 \gamma_2 + t_{34} \gamma_3 \gamma_4 + t_{23}(t) \gamma_2 \gamma_3
\ee
where $\gamma_i$ are Majorana fermion operators, obeying
\be
\gamma_i^\dag =\gamma_i, \ \ \ \ \{ \gamma_i, \gamma_j \} =2 \delta_{ij}
\ee
Any two Majorana fermion operators span a two-state Hilbert space, whose states can be described by a fermion parity $n_f =0,1$.
It is illuminating to re-express the Hamiltonian (\ref{HgEq})  in terms of the  fermion creation and anhiliation operators associated with the fermion parity of each topological wire segment: 
\ba
c^\dag_L =& \frac{1}{2} \left(  \gamma_1 + i \gamma_2 \right) \ \ \ \ c_L =&\frac{1}{2}  \left(  \gamma_1 - i \gamma_2 \right)  \n
c^\dag_R =& \frac{1}{2} \left(  \gamma_5 + i \gamma_6 \right) \ \ \ \ c_R =& \frac{1}{2}  \left(  \gamma_5 - i \gamma_6 \right)
\ea
In the basis of the two fermion parities $n_L$ and $n_R$: 
\be \label{nlrbasis}
|n_L, n_R \rangle =(  |0,0 \rangle, |1,1 \rangle |0,1 \rangle, |1,0 \rangle)^T
\ee
the  Hamiltonian is:
\be \label{HEqn}
H(t) = 
 \begin{pmatrix} t_{12} + t_{34} & t_{23}(t) & 0 & 0 \\ t_{23}(t)  & -t_{12}  - t_{34} & 0 & 0 \\  0 & 0 & t_{12} - t_{34} & t_{23}(t)  \\
0 & 0& t_{23}(t)  & - t_{12} + t_{34}   \\
\end{pmatrix}
\ee

Let us begin by ignoring any coupling between the wire and the outside world, and consider the dynamics implied by the Hamiltonian (\ref{HEqn}).   For the sake of argument let us take $t_{12}, t_{34} >0$, such that for $t_{23} =0$ the ground state of the system in the fermion number basis is  $|n_R, n_L \rangle = |11 \rangle$.  Let us consider begining the experiment with $t_{23}=0$, and with the system in this ground state.  Since $H$ is block-diagonal, we can restrict our attention to the upper block, which is simply a 2-level system analogous to a spin in a magnetic field $\mathbf{B}$.  Here $t_{12} + t_{34}$ plays the role of $B_z$, while $t_{23}$ is analogous to $B_x$.  

While $t_{23} =0$, our proverbial spin is aligned with this ``magnetic field" -- meaning that the system remains in its ground state $|1,1 \rangle$.  Using exactly the same principle as in NMR, we can now turn on  the transverse field $B_x$ (a.k.a $t_{23}$) and make our spin precess about the $x$ axis; if the pulse is turned off after an appropriate time interval the spin will be left in its excited state, anti-aligned with $B_z$.  In other words, the fermion parities will be rotated from  $|1,1 \rangle$ to $|0,0 \rangle$.  
Hence a measurement of the fermion parities immediately following the pulse will find the system in its excited state.  

In a sense this observation seems mundane; it is simply the standard behavior of a quantum 2-level system.  If the levels in question are indeed those of Majorana bound states, however, a positive result demonstrates that turning on $t_{23}$ couples the  Majorana qbit (here the fermion number $n_L$) of the left-hand TSC segment coherently to the Majorana qbit $n_R$ of the right-hand TSC segment.  In other words, a positive result demonstrates that there exist correlations between the two segments that are associated with the non-abelian statstics of the Majorana zero modes.  We will discuss the reasons for this in more detail in Section \ref{MeasurementOnlySec}.
 
How well can our experiment distinguish between Majorana modes and other types of bound states that might arise at the end-points of the topological wire segments?  An important feature of the Majorana system is that tuning the central gate voltage changes the {\it coupling} between the left and right q-bit (in our magnetic analogue, $t_{23}$ is $B_x$).  This is fundamentally different from changing the chemical potential for each 2-state system (in which case $t_{23}$ would be analogous to  $B_z$), since it produces a coherent rotation rather than a transition that occurs by relaxation.  Experimentally the two can be distinguished by verifying that doubling the pulse duration leads to a full precession back to the original state.  Thus the MPR experiment essentially tells us whether the bound states on the two wire segments can be coupled by tuning voltages (or, in alternative scenarios, magnetic fields) in accordance with our expectations for Majorana fermions. 
Unlike a true braiding experiment, however, it  probes a dynamical, non-topological regime, and does not definitively {\it prove} the existence of Majorana statistics; it merely shows that coupling the two wire segments creates a coherent rotation in a 2D subspace of the 4 state Hilbert space, as follows from the non-Abelian statistics if these 4 states arise from Majorana zero modes (see Sect. \ref{MeasurementOnlySec}).  

\subsection{Dynamics in the presence of a bath}

We now turn to a more detailed analysis of the system's dynamics, using the approach described in Ref. \onlinecite{ParityTime}.  What follows is a  discussion of the dynamics of the system when coupled to an external fermionic bath; readers most interested the quantitative outcome of this analysis may skip to section \ref{SomeNumerics}.

In a realistic experimental set-up, the total fermion parity $n_L + n_R$ is not conserved, as the wire cannot be completely isolated from its environment.  To account for this, we couple our Majorana wire to a (non-superconducting) fermionic bath.  The dynamics of the wire system are then described by the Master equation
\be \label{MasterEq}
\dot{\rho} = i \left[ H, \rho \right] + \sum_n \,\Gamma_n\,\left[ L_n \rho L^\dag_n - \frac{1}{2} \left( L^\dag_n L_n \rho + \rho L^\dag_n L_n \right)\right]
\ee
where in the basis (\ref{nlrbasis}) the Linblad operators $L_n$ are given by:
\ba
L_1 = c^\dag_R& =&
 \begin{pmatrix} 0 & 0 &  0 & 0 \\ 
0 &  0 & 0 & -1  \\ 
1 & 0 &0  &0 \\ 
0&0 &0 & 0 \\ 
\end{pmatrix}  \ \ \ \ \ 
L_2 = c_R=
 \begin{pmatrix} 0 & 0 &  1 & 0 \\ 
0 &  0 & 0 & 0  \\ 
0 & 0 &0  &0 \\ 
0& -1 &0 & 0 \\ 
\end{pmatrix}  \n
L_3 = c^\dag_L& =&
 \begin{pmatrix} 0 & 0 &  0 & 0 \\ 
0 &  0 & 1 & 0  \\ 
0 & 0 &0  &0 \\ 
1& 0 &0 & 0 \\ 
\end{pmatrix}  \ \ \ \ \ 
L_4 = c_L =
 \begin{pmatrix} 0 & 0 &  0 & 1 \\ 
0 &  0 & 0 & 0  \\ 
0 & 1 &0  &0 \\ 
0& 0 &0 & 0 \\ 
\end{pmatrix}  \nonumber 
\ea
If the couplings to the bath are homogeneous and the bath itself is in thermal equilibrium with a constant density of states $\rho$, the rates are given by
\ba \label{rates}
\Gamma_1  =& \frac{1}{\hbar}\, |\alpha|^2 \rho \, n_F(-2t_{34}) \ \ \ \ 
\Gamma_2  =& \frac{1}{\hbar}\, |\alpha|^2 \rho \, n_F( 2t_{34})   \n
\Gamma_3  =& \frac{1}{\hbar}\, |\alpha|^2 \rho \,  n_F(-2t_{12}) \ \ \ \ 
\Gamma_4  =& \frac{1}{\hbar}\, |\alpha|^2 \rho \, n_F( 2t_{12})   
\ea
where $\rho$ is the density of states in the bath, $\alpha$ parametrizes the coupling to the bath, and 
\be
n_F (\epsilon)  = \frac{1}{ 1 + e^{ \beta \epsilon}}   
\ee
is the probability that the bath can absorb the energy $ \epsilon$ of the corresponding transition in the wire.

For our Hamiltonian we will of course take (\ref{HEqn}),  with a time-dependent voltage applied to the central gate that causes $t_{23}$ to vary in time according to
\be
t_{23} (t) = \frac{t_{23}^{(0)}}{2} \left [  \theta \left( \cos \omega( t- \delta t/2) \right) - \theta \left( \cos \omega( t+ \delta t/2) \right) \right]
\ee
This gives a square-wave pulse of amplitude $t_{23}^{(0)}$ and duration $\delta t$, with a time lag of $\pi/\omega - \delta t$ between successive pulses.

It is not difficult to simulate the dynamics described by Eq. (\ref{MasterEq}) numerically at arbitrary temperatures and values of the various couplings; we present the results of these simulations in Sect. \ref{SomeNumerics}.  In the limit $T \rightarrow \infty$ (by which in practise we mean that the temperature of the bath is large compared to the couplings $t_{ij}$ between the Majorana fermions -- {\it not} relative to the superconducting gap!), however, the dynamics are sufficiently simple that we can describe them analytically, as we will now do. 

In the high-temperature limit the fermi factors in Eq. (\ref{rates}) are all $1/2$, and there is only one effective relaxation rate $\Gamma \equiv 2 \Gamma_i$.  
The time-dependent density matrix has the general form:
\be
\rho(t) = \begin{pmatrix} \rho_{11}(t) & \rho_{12}(t) & \rho_{13}(t) & \rho_{14}(t) \\
\overline{\rho}_{12}(t) & \rho_{22}(t) & \rho_{23}(t) & \rho_{24}(t) \\
\overline{\rho}_{13}(t) & \overline{\rho}_{23}(t) & \rho_{33}(t) & \rho_{34}(t) \\
\overline{\rho}_{14}(t) & \overline{\rho}_{24}(t) & \overline{\rho}_{34}(t) & \rho_{44}(t) \\
\end{pmatrix}
\ee
where $\rho_{ii}$ are real, and give the probability of finding the system in the corresponding fermion number eigenstate.   Evidently the total probability $\sum_i \rho_{ii}$ must equal $1$.  
The equations of motion are most simply expressed in terms of the quantities
\ba
  \rho_e^{(S)} =\rho_{11} + \rho_{22} \ \ \ \ \    \rho_e^{(A)} =\rho_{11} - \rho_{22} \n
   \rho_o^{(S)} =\rho_{33} + \rho_{44} \ \ \ \ \    \rho_o^{(A)} =\rho_{33} - \rho_{44} \n
 x_{12} = \text{Re} \left( \rho_{12} \right )  \ \ \ \ \ y_{12} = \text{Im} \left( \rho_{12}\right ) \n
  x_{34} = \text{Re}\left(  \rho_{34}\right ) \ \ \ \ \ y_{34} = \text{Im}\left(  \rho_{34} \right )
\ea
Here $\rho_{e}^{(S)}(t)$ ($\rho_{o}^{(S)}(t)$) is the probability of finding the system in a state where the total fermion parity $n_L + n_R$ is even (odd) at time $t$.   $\rho_{e}^{(A)}(t)$ indicates how much more likely we are to find the system in the state $|0,0 \rangle$ than in the state $|1,1 \rangle$; $\rho_{o}^{(A)}$ determines the relative probabilities of measuring the fermion number combinations $|1,0\rangle$ and $|0,1 \rangle$.

The symmetric components $  \rho_e^{(S)},   \rho_o^{(S)}$ decouple from the remaining components; their dynamics are given by
   \ba  \label{Mds}
 \dot{\rho}_e^{(S)}(t)&=& - \Gamma \left( \rho_e^{(S)}(t) -  \rho_o^{(S)}(t) \right ) \n
 \dot{\rho}_o^{(S)} (t)&=&  - \Gamma \left( \rho_o^{(S)}(t) -  \rho_e^{(S)}(t) \right )
 \ea
 This indicates that $ d/dt \text{Tr}( \rho) =0$, and that the probability of finding the system in the sector of even versus odd fermion parity decays exponentially to $1/2$, as it should at high temperature.  

The block off-diagonal components $\rho_{13}, \rho_{14}, \rho_{23}, \rho_{24}$ of the density matrix also decouple from the rest; their dynamics will not be of interest to us here.  

The equations of motion for the remaining six components of $\rho$ reduce to two sets of three coupled first order differential equations:  
\ba
\dot{\rho}_e^{(A)}(t)&=&   - \Gamma \rho_e^{(A)}(t) -4 t_{23}(t)  y_{12}(t) \\
\dot{ x}_{12}(t)&=& - \Gamma x_{12}  + 2  \left( t_{34}+ t_{12}  \right)y_{12}(t) \n
\dot{y}_{12}(t)&=&-  \Gamma y_{12}  - 2  \left( t_{34}+ t_{12} \right)x_{12}(t) + t_{23}(t) \rho_e^{(A)} \nonumber 
\ea

\ba 
 \dot{\rho}_o^{(A)}(t)&=& - \Gamma \rho_o^{(A)}(t)   - 4  t_{23}(t) y_{34}(t) \\   
\dot{x}_{34}(t)&=& - \Gamma x_{34}  - 2  \left( t_{34}- t_{12} \right)y_{34}(t) \n
\dot{ y}_{34}(t)&=&-  \Gamma y_{34} + 2  \left( t_{34}- t_{12}  \right)x_{34}(t) + t_{23} (t)\rho_o^{(A)} \nonumber 
    \ea
    where we have temporarily set $\hbar =1$ for notational simplicity.

For $t_{23}$ independent of time, each set can be solved analytically to give:
\begin{widetext}
\ba \label{Tsols1}
\rho_e^{(A)} (t)\ &=&  \frac{ e^{-\Gamma t} }{\epsilon_e^2}\left(\rho_e^{(A)}(0)\left(
   t_{23}^2 \cos (2 \epsilon_e t)+ t_{12,34}^2\right)+2 x_{12}(0) t_{23}  t_{12,34}  \left( 1 -\cos (2 \epsilon_e t) \right)
   -2 y_{12}(0) \epsilon_e t_{23} \sin (2 \epsilon_e t) \right )
  \n
 x_{12}(t)&=& \frac{e^{-\Gamma t} }{2  \epsilon_e^2}
   \left( 2 x_{12}(0) \left(  t_{23}^2 - t_{12,34} ^2\cos (2 \epsilon_e t) \right ) +\rho_e^{(A)}(0)
   t_{23} t_{12,34} \left(  1 - \cos (2 \epsilon_e t) \right)+2 y_{12}(0) \epsilon_e t_{12,34} \sin (2 \epsilon_e
   t)  \right)\n
y_{12}(t)&=& \frac{e^{-\Gamma t} }{2 \epsilon_e}\left ( \rho_e^{(A)}(0) t_{23} \sin (2 \epsilon_e t)-2 x_{12}(0)
   t_{12,34} \sin (2 \epsilon_e t) +2 y_{12}(0) \epsilon_e \cos (2
   \epsilon_e t)\right)
\ea
\end{widetext}
where $t_{12,34} = t_{12} + t_{34}$, and $\epsilon_e = \sqrt{ (t_{12} + t_{34})^2 + t_{23}^2 }$ is the modulus of the band energy in the even-parity sector.   (The form of the solution is similar in the odd sector).

We are interested in the behavior of the diagonal component $\rho_e^{(A)}(t)$.  For simplicity, let us begin in the ground state for $t_{23} =0$, for which the density matrix has only one non-vanishing component $\rho_{22}$.   For $t_{23} =0$, the equations of motion are simply
\ba \label{}
\rho_e^{(A)} (t)\ &=&   e^{-\Gamma t}  \rho_e^{(A)}(0) 
    \n
 x_{12}(t)&=& 
y_{12}(t) =0
\ea
such that the system relaxes at a rate $\Gamma $ towards the equilibrium density matrix $\rho_{ii} = 1/4$.  If we turn on $t_{23}$ at time $t= t_0$, however, the new equations of motion are:
\ba
\rho_e^{(A)} (t)\ &=&  \frac{ e^{-\Gamma t } }{\epsilon_e^2}\left(\rho_e^{(A)}(0)\left(
   t_{23}^2 \cos (2 \epsilon_e (t- t_0))+ t_{12,34}^2\right)   \right )
  \n
 x_{12}(t)&=& \frac{e^{-\Gamma t} }{2  \epsilon_e^2}
   \left( \rho_e^{(A)}(0)
   t_{23} t_{12,34} \left(  1 - \cos (2 \epsilon_e (t- t_0)) \right) \right)\n
y_{12}(t)&=& \frac{e^{-\Gamma t} }{2 \epsilon_e}\left ( -\rho_e^{(A)}(0) t_{23} \sin (2 \epsilon_e (t- t_0)) \right)
\ea
If we return $t_{23}$ to $0$ after a time $\delta t = \pi/(2  \sqrt{ (t_{12,34})^2 + t_{23}^2 })$, then
$\rho_e^{(A)}$ evolves from its initial value of $e^{-\Gamma t_0}$ to a minimal value of 
\be \label{RhominVal}
\rho_e^{(A)} (t_0 + \delta t) = e^{-\Gamma (t _0 + \delta t) } \frac{ t_{12,34}^2 - t_{23}^2}{t_{12,34}^2+t_{23}^2}\rho_e^{(A)}(0)
\ee
At this point $x_{12} = t_{23} t_{12,34}/ \epsilon_e^2, y_{12} =0$; the ensuing dynamics is given by substituting these initial values into Eq. (\ref{Tsols1}) with $t_{23} =0$.   Note that once we set $t_{23}$ to 0 we will not see oscillations in $\rho_e^{(A)}$, though the values of  $x_{12}$ and $y_{12}$ will oscillate.  If the delay between applications of $t_{23}$ is an integer multiple of $2 \pi/(t_{12}+ t_{34})$, then at the beginning of the second pulse we once again have $x_{12}= t_{23} t_{12,34}/ \epsilon_e^2, y_{12} =0$.  In this case, on the second pulse, $\rho_e^{(A)}$ starts at its minimal value
$e^{-\Gamma t } \frac{ t_{12,34}^2 - t_{23}^2}{t_{12,34}^2+t_{23}^2}$,
and after a time $\pi/(2 \epsilon_e)$ has returned to a maximal value of $e^{-\Gamma( t + \delta t)} $.  

Note that the maximum contrast is obtained when $t_{23}$ is pulsed to a value that is large compared to $t_{12} + t_{34}$, which corresponds to applying a large $B_x$ field.  In this case, in the limit $\Gamma \rightarrow 0$, subsequent pulses cause $\rho_e^{(A)}$ to oscillate between $-1$ and $1$.

\subsection{Finite voltage ramp time} \label{VoltageSect}

In practise, $t_{23}$ cannot be tuned from $0$ to $t_{23}^{(0)}$ instantaneously.  In our numerical simulations, we therefore include the effect of a finite voltage ramp time, via 
\ba
t_{23} (t) &=& \frac{1}{\pi} \left | \tan^{-1} \left[ s \cos\left( \omega (t - \delta t/2) + \phi \right)  \right ]  \right .\n
&& \left. - \tan^{-1} \left[ s \cos\left( \omega (t + \delta t/2)+ \phi \right) \right ] \right |
\ea
where $\phi$ determines the time at which the first pulse is applied, $\delta t$ is the duration of the pulse, $\omega$ determines the time between pulses, and $s$ determines the voltage ramp time, which we typically take to be of order $1  \, ns$ unless otherwise specified.  Note that in order for the signal to be observable, the voltage ramp time cannot be too long: it is crucial that $t_{23}$ is turned on {\it non}- adiabatically.  

\section{Numerical Results} \label{SomeNumerics}

Let us now turn to the question of what the dynamics described in the previous section implies for our experiment.  
We will assume that we can prepare the system in the state $|1,1 \rangle$, either by allowing relaxation to the ground state (in the limit that the bath is at low temperature), or by an explicit measurement of the fermion parity
 (if the bath is at high temperature).  In this case the signal that we are interested in is the probability of finding the system in the state $|0,0 \rangle$ at the end of the sweep, indicating that the qbits in the two TSC wires have been coherently coupled.  
This signal can be read off from the time dependence of the diagonal components of the density matrix, which give the probabilities of the four possible measurement outcomes.   We will use the following convention when plotting these: the probability of finding the system in the state $|n_L,n_R \rangle = |0,0 \rangle$ is $\rho_{11}$, which will be shown in green.  Similarly the probability of finding  state $|1,1 \rangle$ is $\rho_{22}$ (red); state $|0,1 \rangle$ is $\rho_{33}$ (purple), and  state $|1,0 \rangle$ is $\rho_{44}$ (blue).

\subsection{Experimental parameters}

There are three free parameters that are important in determining the system's behaviour.  First, as is apparent from Eq. (\ref{RhominVal}), the minimum value of $\rho_{22} - \rho_{11}$ (which gives the difference in the probabilities of finding the system in the state $|1,1 \rangle$ or in the state $|0,0 \rangle$) is determined by the relative magnitudes of $t_{23}^{(0)}$ and $t_{12} + t_{34}$.  Hence the greatest signal will be observed if the maximum value of $t_{23}$ is large compared to the couplings $t_{12}, t_{34}$ between the Majorana zero modes on the same TSC segment.  Second, the ratio of the relaxation rate $\Gamma$ to the scale of the couplings $t_{ij}/\hbar$ imposes an absolute limit on the magnitude of the signal, even for $t_{23}^{(0)} \gg t_{12} + t_{34}$; if the relaxation rate is too high the system will reach equilibrium before the pulse can be completed.  Third, the temperature $T_{\text{bath}}$ of the bath relative to $|t_{12} + t_{34}|$ determines which relaxation processes can occur.  If the bath temperature is high on this scale, then as discussed in the previous section relaxation occurs for all values of $t_{23}$, meaning that the relaxation time must be longer than the total time required to initialize the system, apply the pulse, and take a final measurement.  If the bath temperature is low on this scale, on the other hand, relaxation occurs only while the system is in its excited state; initialization is not required, and the limiting factor is the ratio of the relaxation time of the excited states to the duration of the pulse.  

Appropriate values for these parameters can be estimated from the literature.  
Let us suppose that each of the TSC segments are approximately $1 \mu m$ long.  In this case, we expect that by tuning either the chemical potential or Zeeman field, the splitting between the two fermion parity states in each segment is expected to be in the range of approximately $t_{12}, t_{34} \approx 0.5$ to $30 \ \mu eV$\cite{RainisPRBB87,DasSarmaSmoking}.  By adjusting the length of the  NSC segment, we expect that $t_{23}$ can be tuned along a comparable range of values.  (Note, however, that the oscillations as a function of $\mu$ and the Zeeman field expected in $t_{12}$ and $t_{34}$ are often absent in $t_{23}$, as explained in Appendix \ref{WaveFnApp}, such that the voltage pulse applied to the central gate must be relatively large).  
This gives a pulse duration in the approximate range of $0.05$ to $2 \ ns$.  At these values current experiments, which are performed in the range of $50-75 \ mK \sim 4 -6 \ \mu eV$, can be in the relatively low-temperature or relatively high-temperature regime, depending on whether the $t_{ij}$ are at the high end or the low end of the expected possible range.  (In practise, the distribution of quasi-particles in the bulk superconductor is generally not thermal\cite{RainisLossPRB85.174533}, such that the effective temperature may be considerably higher than this due to quasiparticle poisoning).  

A pessimistic estimate of the relaxation rate was obtained by Ref. \onlinecite{RainisLossPRB85.174533}, who suggested that in nanowire systems of the type considered here $\Gamma^{-1} \approx 10^{-7} s$.  This would be problematic for applications to adiabatic quantum computing, but is nonetheless two orders of magnitude longer than the pulse duration required to perforrm the coherent rotation, suggesting that signatures of Majorana correlations can in principle be detected.

The parameter values used for these estimates, and the important time-scales for the experiment, are summarized in Table \ref{NumbersTab}.  

\begin{table}[h!]
\begin{tabular}{|c|c|}
\hline 
Parameter & approximate value   \\
\hline
$T$ & $4-6 \ \mu eV$ \\
$t_{12}, t_{34} $& $0.5- 30 \ \mu eV$ \\
$t_{23}$ & $0.5 - 30 \ \mu eV$ \\
$\delta t $ &$ \stackrel{<}{\approx} 2 \ ns $ \\
$\Gamma^{-1}$ & $ <100 \ ns$\\
\hline
\end{tabular}
\caption{ \label{NumbersTab} Approximate range of parameter values, based on the experimental set-up of Ref. \onlinecite{MourikScience336}, and parameter estimates of Refs. \onlinecite{RainisPRBB87,DasSarmaSmoking,RainisLossPRB85.174533} for a wire with TSC segments approximately $1 \ \mu m$ long.}  
\end{table}

\subsection{Dynamics}

\begin{figure}
\begin{tabular}{ll}
 (a)&  \\
 & \includegraphics[width=0.8\linewidth]{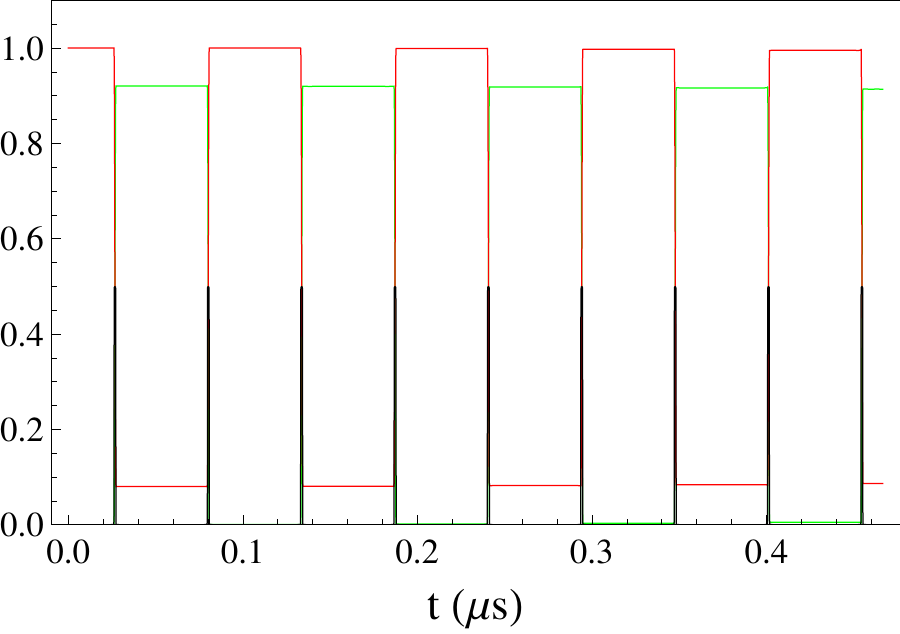}  \\
   (b)& \\  
   & \includegraphics[width=0.8\linewidth]{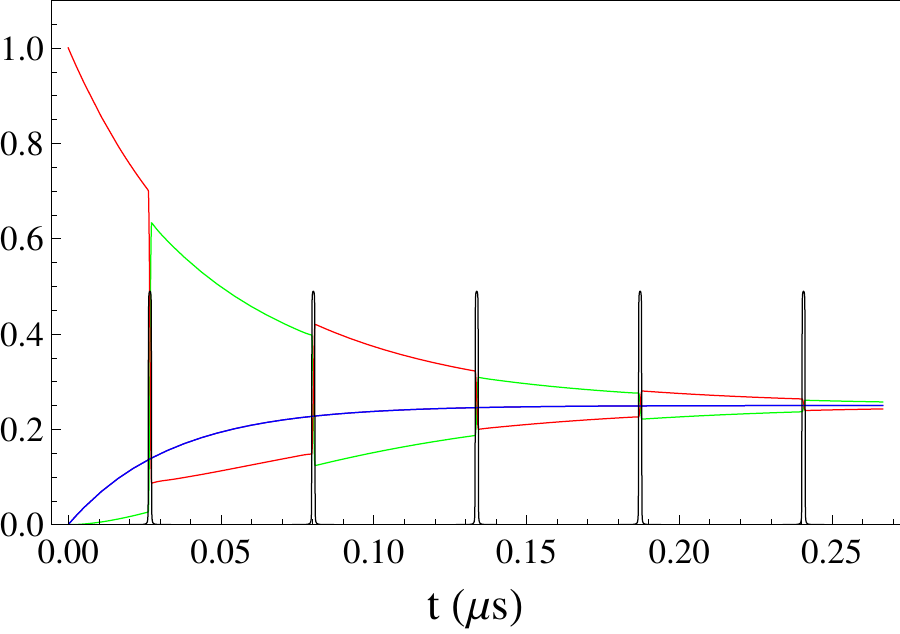} \\ 
   \end{tabular}
 \caption{Time evolution of the diagonal elements of the density matrix in the limit of extremely slow relaxation rates.  $\rho_{11}$ is shown in green;  $\rho_{22}$ is red;  $\rho_{33}$ is purple, and  $\rho_{44}$ is blue.  We initialize the system in a state with fermion parity $|1,1 \rangle$ at time $t= 0$, and after some time apply a voltage pulse (indicated by the black line) to the central gate that switches $t_{23}$ from $0$ to $t_{23}^{(0)}$.  Both plots are shown for the choice $t_{23}^{(0)}/(2 \pi \hbar) =232 \, MHz, t_{12}/(2 \pi \hbar)  = t_{34}/(2 \pi \hbar)  =34 \, MHz$, for which $\epsilon_e/(2 \pi \hbar) =242 \, MHz$, or $\epsilon_e = 1 \, \mu eV$.
 (a) For $\Gamma= 0$ the pulse rotates the system from the ground state $|1,1 \rangle$ to a state that is predominantly composed of the excited state $|0,0 \rangle$.  A second pulse after an appropriate delay returns the system to the ground state.
 (b) For $\Gamma = 15 \, MHz$, 
 the oscillation decays exponentially over several periods.  The two states $|1,0 \rangle $ and $|0,1\rangle$ in the odd fermion parity sector are generated with equal probability by relaxation processes.}
\label{LowGamPlots}
 \end{figure}

We now discuss the  qualitative nature of the dynamics.  
Figure \ref{LowGamPlots} shows the situation for long relaxation times, which we expect to be relevant for experiments.  If $\Gamma=0$ (Fig. \ref{LowGamPlots} a), the system remains in the sector of even fermion parity indefinitely, and oscillates between being found in the ground state $|1,1 \rangle$ and the excited state $|0,0\rangle$ as subsequent pulses in $t_{23}$ are applied.  Tuning the pulse width and inter-pulse delay appropriately results in a stable period 2 oscillation, as shown here.  With a more realistic value $\Gamma = 15 \, MHz$ (Fig. \ref{LowGamPlots} b), at high bath temperatures this oscillation is damped over many cycles.  
Because we have initialized the system in the sector of even fermion parity, the states $|1,0 \rangle$ and $|0,1 \rangle$ are produced only by relaxation processes.  For the choice $t_{12} = t_{34}$ (as is the case in the plots shown here) the two parity odd states are degenerate, and $\rho_{33} = \rho_{44}$ irrespective of $T_{\text{bath}}$.

\begin{figure}
\includegraphics[width=0.8\linewidth]{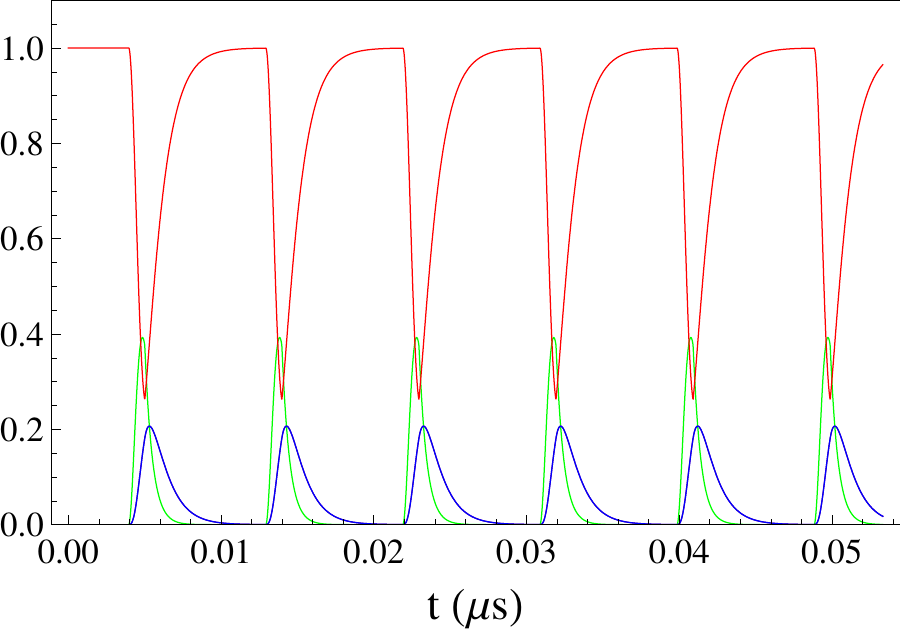}  
 \caption{Time evolution of the diagonal elements of the density matrix for low bath temperature and high relaxation rates, shown here for  $t_{23}^{(0)}/(2 \pi \hbar) =232 \, MHz, t_{12}/(2 \pi \hbar)  = t_{34}/(2 \pi \hbar)  =34 \, MHz$, and an effective relaxation rate of $\Gamma=1 \, GHz$ from the excited states.  Relaxation occurs only when the system is in an excited state, meaning that the system starts to relax when the pulse in $t_{23}$ is applied.  After the end of the pulse the system relaxes to its ground state  $|1,1 \rangle$, effectively re-initializing the process. }
\label{LowTPlots}
 \end{figure}

It is also interesting to consider the case where $\hbar \Gamma$ is of the same order of magnitude as the couplings between the Majorana fermions.  This could occur, for example, if the parity measurement has a strong impact on relaxation rates.  
%
%
%
In the limit that the bath temperature $T_{\text{bath}}$ is small on the scale of $t_{ij}/\hbar$, the result is shown in Fig. \ref{LowTPlots}.  
At low temperature relaxation processes only begin when $t_{23}$ is switched on, and   
the lower bound on the relaxation time $\Gamma^{-1}$ for the signal to be visible is that it must not be too much shorter than the duration of the pulse, $\delta t \approx \pi \hbar /(2 \sqrt{ (t_{12}+t_{34})^2 + t_{23}^2 } )$.\footnote{In our analytical discussion of $\delta t$, we did not account for the effect of the finite time required to change $t_{23}$.  In our simulations we take this time to be small but finite, resulting in a small correction to the idealized value of $t_{23}$.}    Since the system relaxes to its ground state after the pulse is completed, each subsequent pulse begins in the same initial state $|1,1 \rangle$.  Hence repeated pulses produce persistent oscillations in the probabilities of finding the system in each of the 4 possible states.
If the time required to make a measurement is fast compared to the pulse duration, we can obtain similar results at high bath temperatures by performing a parity measurement immediately before each pulse, which  initializes the system into a state of definite fermion parity. 
(Note that if a measurement  initializes the system in 
either of the states $|1,0 \rangle$ or $|0,1 \rangle$, the ideal pulse duration $\delta t_{o} \approx \pi   \hbar/(2 \sqrt{ (t_{12}-t_{34})^2 + t_{23}^2 } )$ is different and the signal will be suppressed.)

\begin{figure}
\includegraphics[width=1.0\linewidth]{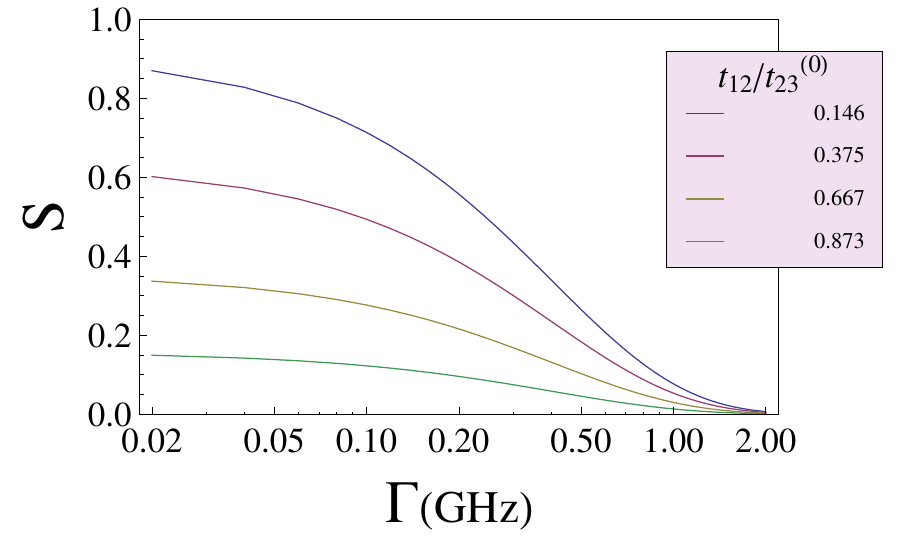}  
 \caption{The magnitude of the signal $S$ (Eq. \ref{ContrastEq}) as a function of the relaxation rate $\Gamma$
 for $K_B T_{\text{bath}} \gg |t_{12} + t_{34}|, t_{23}^{(0)} $, shown for $t_{12} = t_{34}$, $\epsilon_e/(2 \pi \hbar) =   242 \, MHz$ (corresponding to $\epsilon_e = 1 \, \mu eV$), and for several choices of the ratio $t_{12}/t_{23}^{(0)}$.  The $x$ axis here is shown on a log scale, to better display the contrast for the relatively small values of $\Gamma$ expected to be relevant to experiments. 
}
\label{ContrastPlot}
 \end{figure}

Hence if the initialization and measurements themselves can be carried out relatively quickly, even at relaxation rates that are very high compared to the estimate of Ref. \onlinecite{RainisLossPRB85.174533}, our experiment should be able to detect Majorana correlations provided that  the duration of the pulse, $\delta t = \hbar \pi/(2  \sqrt{ (t_{12}+t_{34})^2 + t_{23}^2 } )$ is not long relative to the relaxation time $\Gamma^{-1}$.  Essentially, this implies that the energy scale for the coupling to the bath must not be large compared to the energy scale of the couplings between the Majorana zero modes.  As a rough benchmark, let $\rho_e^{(A)} \equiv \rho_{11} - \rho_{22}$, and imagine that we initialize the system at time $t_0$ such that $\rho_{e}^{(A)} =-1$, meaning that the system is in its ground state $|1,1 \rangle$.  We define the magnitude of the signal to be the {\it difference} between the value of $\rho_e^{(A)}$ at time $t_0 + \delta t$ with and without the pulse in $t_{23}$:
 \be \label{ContrastEq}
S = \frac{1}{2} \left( \rho_e^{(A)} (\delta t)|_{t_{23} = t_{23}^{(0)} } -  \rho_e^{(A)} (\delta t)|_{t_{23} = 0 } \right) \ .
\ee
The maximum possible value of $S$ is $1$, which is attained in the limit $t_{23}^{(0)} \gg |t_{12} + t_{34}|$, $\Gamma =0$.  The minimum value is $0$, which indicates that the system has relaxed completely before the pulse has been completed.  
Fig. \ref{ContrastPlot} shows the magnitude of $S$ for $t_{23}^{(0)}/(2 \pi \hbar) =232 \, MHz, t_{12}/(2 \pi \hbar)  = t_{34}/(2 \pi \hbar)  =34 \, MHz$, as a function of the relaxation rate $\Gamma$, in the limit $K_B T_{\text{bath}} \gg |t_{12} + t_{34}|,  t_{23}^{(0)}$.

\subsection{Effects of finite voltage ramp time}

In practise, at the frequencies that we are interested in, we must contend with the finite response time required to charge the gate.  
We model these numerically as described in Sect. \ref{VoltageSect}.  
Here we discuss the quantitative importance of the voltage ramp time.  

Intuitively, the limiting factor is that if the system is to be found in the excited state after the pulse, then the charging time of the gate must not be adiabatic at the energy scale of the Majorana Hamiltonian.  Indeed the closer the charging time is to the adiabatic limit, the smaller the probability of finding the system in the excited state immediately after the pulse.  With the parameter values used here, the energy splittings of the Majorana Hamiltonian vary between approximately $0.2$ and $2 \, \mu eV$, which means that $t_{23}$ must ramp to its maximum value on a time-scale that is fast compared to $\sim 250 MHz$.  In practise his requires a charging time of at most a few $ n s$ in order for the a significant signal to be observed.  

%

%

The change in voltage required to tune $t_{23}$ appropriately is discussed in Appendix \ref{WaveFnApp}.  In principle it is possible to tune the couplings $t_{12}, t_{34}$ to be arbitrarily small, requiring smaller maximal values of $t_{23}^{(0)}$ to perform the rotation, as well as implying longer pulse times and allowing the experiment to be performed at longer gate-charging times.

\section{The MPR experiment and braiding via measurement} \label{MeasurementOnlySec}

In the previous section, we showed that pulsing the coupling between the two Majorana fermions $\gamma_2, \gamma_3$ in Fig. \ref{WireFig} produces a coherent rotation of the fermion parities of the two TSC segments.  
Here we will discuss in more depth the connection between this coherent rotation and the non-abelian statistics of Majorana zero modes.

\begin{figure}
 \includegraphics[width=1.0\linewidth]{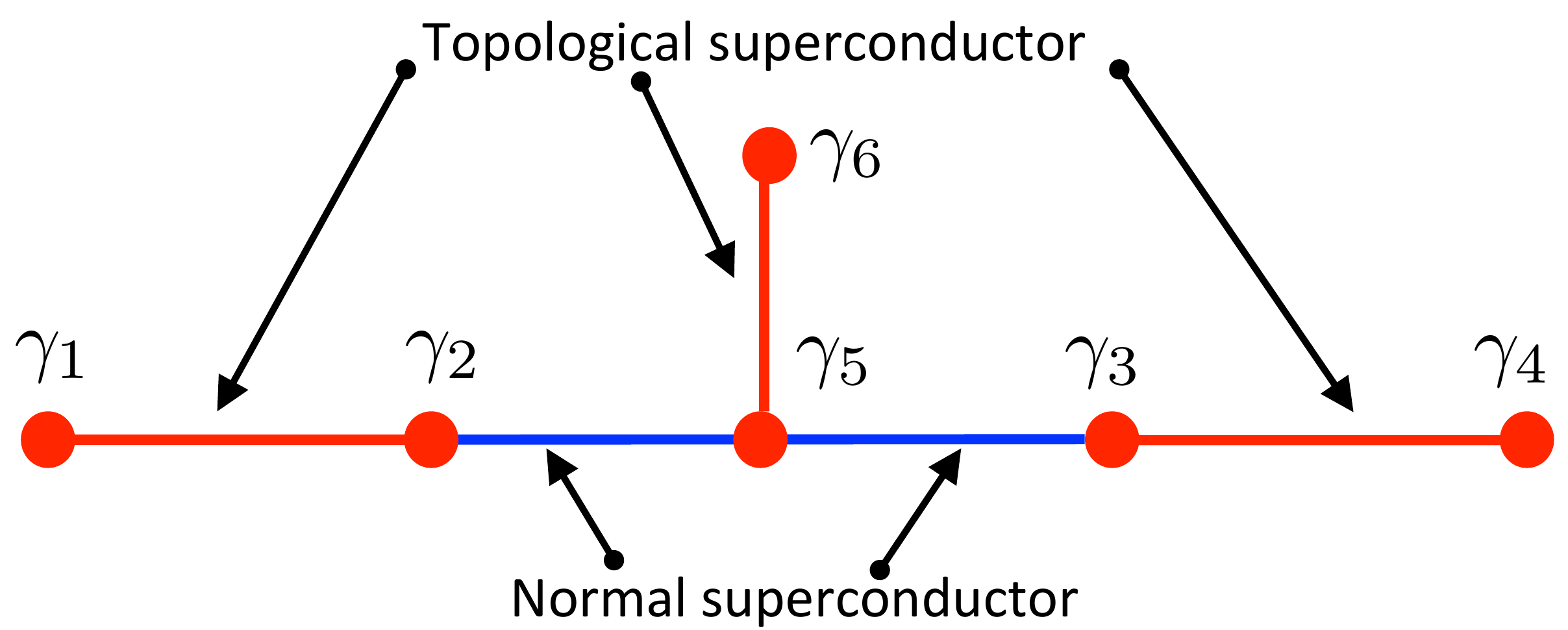} 
 \caption{ Experimental set-up for 1D superconducting wires required for the projective braiding experiment.  Red lines indicate segments of the wire that are in the topological superconducting state; blue lies indicate segments in the normal superconducting state.  The red dot at the boundary between normal and topological superconducting regions indicates the candidate Majorana zero mode. }
\label{PBRaidExp}
 \end{figure}

To understand this connection, we first observe that 
the MPR experiment is closely related to another
 experiment, which we call the projective braiding experiment.  The set-up for this experiment (Fig. \ref{PBRaidExp}) requires two wires, joined at a `tri-junction' by a segment of normal superconducting wire.  The horizontal wire has two topological superconducting segments, joined by a normal superconducting segment; mid-way through this normal segment there is a tri-junction with a vertical TSC wire.   The red dots in the Figure indicate the six Majorana fermions localized at the end-points of the three TSC segments.
 
 The projective braiding experiment is performed in the topological regime, where the couplings between all pairs of Majoranas are tiny.  It implements the ``measurement only" approach\cite{BondersonMeasurement} to exchanging two Majorana zero modes.  Specifically, we will exchange $\gamma_2$ and $\gamma_3$ in Fig. \ref{PBRaidExp} by performing a sequence of measurements of the fermion parity in the TSC wire system.  This is equivalent\cite{BondersonPRB035113} to other 
 proposals\cite{AliceaNatPhys7,SauPRB84} for using tri-junctions to exchange these anyons; such an experiment could therefore be used to detect the non-Abelian statistics of the Majorana zero modes.

\begin{figure}
 \includegraphics[width=0.5\linewidth]{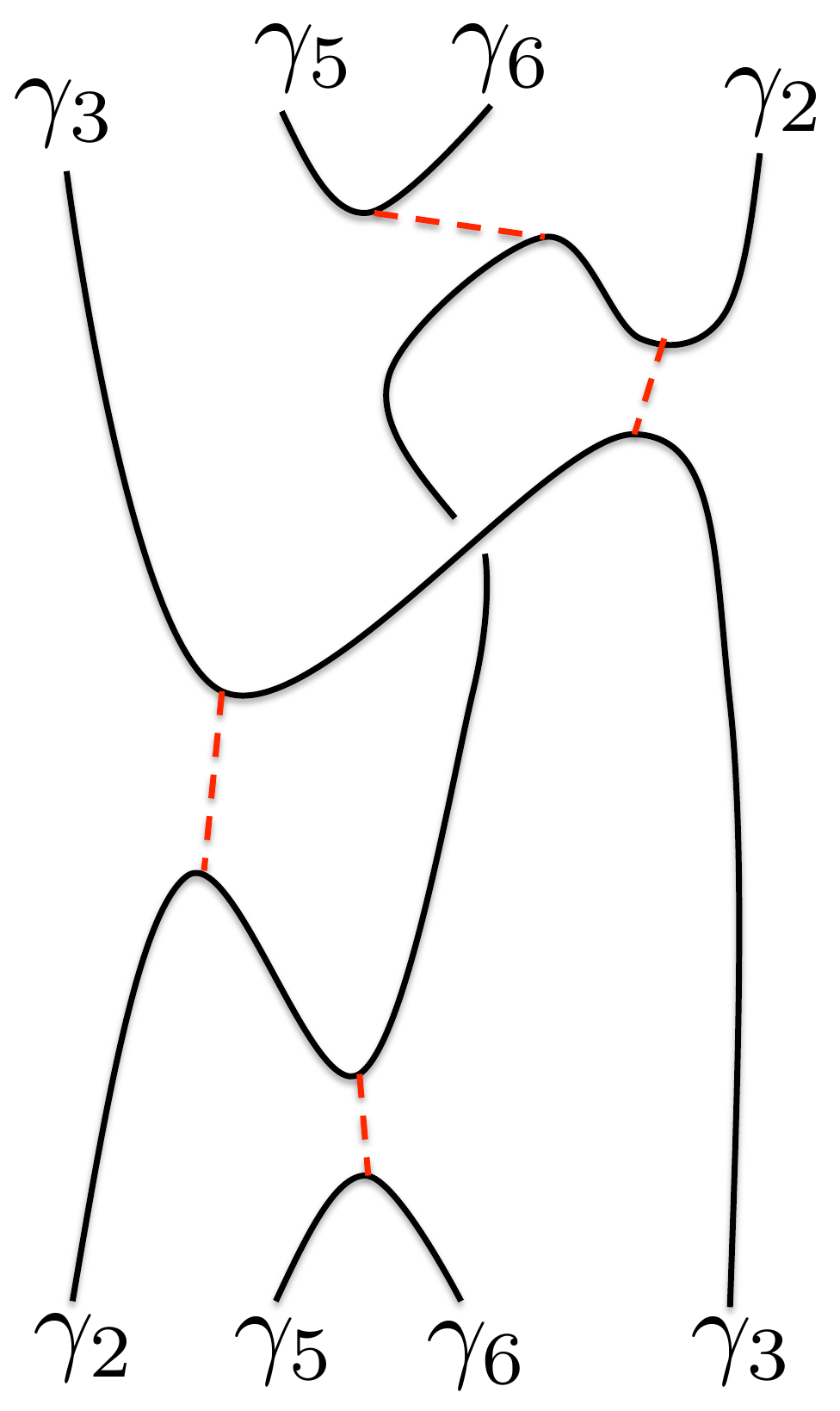} 
 \caption{ Exchanging $\gamma_2$ and $\gamma_3$ through the sequence of projections described in the text\cite{BondersonMeasurement}.  Solid lack lines represent the Majorana zero modes; dashed red lines represent projections onto the state $n_f^{(ij)} =0$. }
\label{MeasurementBraids}
 \end{figure}

We will begin by explaining the principle of the projective braiding experiment\cite{BondersonMeasurement}.   Recall that any two Majorana zero modes $\gamma_i, \gamma_j$ comprise a two-state Hilbert space, which can be described by a fermion number $n_f ^{(ij)}=0,1$. The measurement-only protocol relies on the fact that by repeated measurements, one can always project onto the subspace $n_f^{(ij)} =0$, and that such projections can be used to carry out braiding operations.  Specifically, as proven in Ref. \onlinecite{BondersonMeasurement}, the following sequence of projections effectively exchanges the Majorana fermions $\gamma_2$ and $\gamma_3$:
(1) Project $(\gamma_5, \gamma_6)$ onto $n_f^{(56)} =0$; \\
(2) project $(\gamma_2, \gamma_5)$ onto $n_f^{(25)}=0$; \\
(3) project $(\gamma_5, \gamma_3)$ onto $n_f^{(53)} =0$; \\
(4)project $(\gamma_5, \gamma_6)$ onto $n_f^{(56)} =0$.\\
 Fig. \ref{MeasurementBraids} illustrates how this effectively exchanges the two Majorana zero modes.

Now, let us understand how this works in our wire system in practise.  
To do so, 
it is convenient to re-express these operations in terms of the fermion creation and anhiliation operators associated with the fermion parity of each topological wire segment: 
\ba \label{TheFermionBasis}
c^\dag_L =& \frac{1}{2} \left(  \gamma_1 + i \gamma_2 \right) \ \ \ \ c_L =&\frac{1}{2}  \left(  \gamma_1 - i \gamma_2 \right)  \n
c^\dag_R =& \frac{1}{2} \left(  \gamma_3 + i \gamma_4 \right) \ \ \ \ c_R =& \frac{1}{2} \left(  \gamma_3 - i \gamma_4 \right)  \n
c^\dag_M =& \frac{1}{2} \left(  \gamma_5 + i \gamma_6 \right) \ \ \ \ c_M =& \frac{1}{2}  \left(  \gamma_5 - i \gamma_6 \right) 
\ea

The projectors used to carry out the experiment are:
\be
P_{n_f^{(ij)} = 0} =(1 - n_f^{(ij)}) =  \frac{1}{2} \left( 1 + i \gamma_i \gamma_j \right )
\ee
In terms of the fermion operators defined in Eq. (\ref{TheFermionBasis}), $P_{n_f^{(56)} =0} = 1- n_M$, and the remaining projection operators are:
 \ba
\frac{1}{2} \left( 1 + i  \gamma_2 \gamma_5 \right ) =&\frac{1}{2} \left[ 1 +   \left(  c^\dag_L c^\dag_M + c_M c_L +c^\dag_L c_M +  c^\dag_M c_L \right ) \right] \n
\frac{1}{2} \left( 1 +i  \gamma_5 \gamma_3 \right)=& \frac{1}{2} \left[ 1 - i \left(  c^\dag_R c^\dag_M - c_M c_R +c^\dag_R c_M -  c^\dag_M c_R \right )\right] \n
\ea
Performing steps (2) and (3) amounts to acting with the product operator:
\be \label{23Op}
P_{n_f^{(35)} = 0} P_{n_f^{(25)} = 0} = \frac{1}{4} \left( 1 +i  \gamma_5 \gamma_3 \right) \left( 1 + i  \gamma_2 \gamma_5 \right ) 
\ee
It is useful to separate the terms in this product into two types:  terms with an odd number of $\gamma_5\equiv c^\dag_M + c_M$ operators, which change the eigenvalue of $n_M$, and terms with an even number of $\gamma_5 $, which do not.  The former will be anhiliated when the operator (\ref{23Op}) is sandwiched between two applications of $ (1- n_M)$.  Hence the operator that performs the exchange-via-measurement operation is
\ba
\frac{1}{4} P_{n_f^{(56)} = 0} \left( 1 + i \gamma_5 \gamma_3 \right ) \left( 1 + i \gamma_2 \gamma_5 \right ) P_{n_f^{(56)} = 0} \n
=\frac{1}{4}  P_{n_f^{(56)} = 0} \left( 1 + \gamma_2 \gamma_3 \right ) P_{n_f^{(56)} = 0} 
\ea

Since $\gamma_5, \gamma_6$ enter into this expression only in the initial and final projectors, we can restrict our attention to the subspace of the total Hilbert space in which $n_f^{(56)} =0$.  In other words, we can remove the vertical wire from Fig. \ref{PBRaidExp} entirely, replacing the 2-level system $\gamma_5, \gamma_6$ of the vertical TSC wire with the vacuum state, since this is the only state that enters into the exchange process.  
This results in the single-wire geometry of Fig. \ref{WireFig}, with the exchange process being carried out by the hopping operator $\frac{1}{4}(1 + \gamma_2 \gamma_3)$.  

Perhaps the clearest signature of the non-Abelian statistics of Majorana zero modes is the following.  Consider two pairs of Majorana zero modes, a left pair with fermion parity $n_L =\frac{1}{2} ( 1 - i \gamma_1 \gamma_2)$ and a right pair with fermion parity $n_R=\frac{1}{2} ( 1 - i \gamma_3 \gamma_4)$.  Let us begin in the state $n_L = n_R =0$.  We then exchange the $\gamma_2$ and $\gamma_3$ particles twice (aka braiding $\gamma_2$ around $\gamma_3$).  After this braiding process, the system will be in the state $n_L = n_R =1$.  Performing the same operation again returns the system to its original state $|0,0 \rangle$.  In other words, the braiding process coherently rotates $|n_L, n_R \rangle = |0,0 \rangle$ into $|n_L, n_R \rangle = |1,1 \rangle$, and vice versa.

Such a braiding operation is carried out by acting twice with the exchange operator, which gives:
\be
\frac{1}{16}(1 + \gamma_2 \gamma_3)^2 = \frac{1}{8} \gamma_2 \gamma_3
\ee
This is just the hopping term with coefficient $t_{23}$ in Eq. (\ref{HgEq}).

This allows us to connect the results of the previous sections to the non-Abelian statistics of the Majorana zero modes: by turning on the coupling $t_{23}$ (and simultaneously turning off the other couplings $t_{12}, t_{34}$) for an appropriate amount of time, we are effectively carrying out a braiding operation of the two Majorana zero modes $\gamma_2$ and $\gamma_3$.  The non-abelian statistics dictate that this performs a coherent rotation of the ground state $|1,1 \rangle$ into the excited state $|0,0 \rangle$.
From the topological point of view, this is the underlying reason that our experiment works.

\section{Conclusions} 

In this work, we describe a relatively simple experiment which uses a simultaneous measurement of the fermion parity in two segments of a topological superconducting wire to infer correlations that give a strong indication of underlying non-abelian statistics.  Our proposed experiment uses an approach very similar in spirit to NMR, in which the system is prepared in its ground state, and coherently rotated into an excited state by applying a ``transverse field" (in this case, by changing a gate voltage under one segment of the wire) for a specified amount of time.  The fact that the system is in its excited state after applying the pulse is directly related to the non-abelian statistics of the Majorana zero modes.  It also illustrates that there are non-topological ways to coherently manipulate pairs of Majorana q-bits (which are not truly topological unless the couplings between them vanish) -- a fact that must be carefully taken into account in any implementation of truly topological quantum information processing.

Experimentally this approach presents several challenges.  First, to maximize the signal one would like to be able to tune the coupling between the two Majoranas $\gamma_2$ and $\gamma_3$ separated by the non-topological wire segment (see Fig. \ref{WireFig}) between a value that is very small, and a value that is larger than the couplings between the Majoranas on the same TSC segment.  
For Majorana splittings $t_{ij}$ on the order of $\mu eV$, this requires tuning the voltage on the central gate on a nanosecond time-scale.  In the alternative geometry in which this coupling is tuned by a magnetic flux, it is the flux that must be tunable at this rate.  
Second, the experiment requires a measurement of the fermion parity of the wire segment, which is also a significant experimental challenge. 

However, dynamical experiments of the type described here are nonetheless much easier to carry out than  adiabatic braiding experiments of the type discussed in \cite{AliceaNatPhys7,SauPRB84}.  In our setup, a signal can be obtained even at relatively unfavourable values of the lifetime, and even if the bath temperature is not small compared to the couplings between the Majorana zero modes.  Indeed the limiting factor in our experiment is that the Majoranas must not be more strongly coupled to an external bath than they are to each other -- a limit that is expected to be comfortably satisfied by current experiments.  Hence we expect that dynamical experiments of the type described here will give the first evidence of non-Abelian statistics in these systems.


{\bf Acknowledgements}  FJB wishes to thank Yuval Oreg, Alexander Shnirman, E. Burnell and Steve Simon for useful conversations, and acknowledges KITP (NSF Grant no. PHY11-25915) for its hospitality.

\appendix

\section{Evaluating $t_{ij}$ across TSC and NTSC wire segments}
\label{WaveFnApp}

For the purposes of our experiment, it is important to have an accurate estimate of the $t_{ij}$ in Eq. \ref{HgEq}.  Here we will briefly review how these are calculated.  We follow closely the method described in Ref. \onlinecite{DasSarmaSmoking}, applied to the slightly different geometry shown in Fig. \ref{WireFig}.  

The Hamiltonian along the wire is of the form
\be \label{HSC}
H= \left( - \partial^2_x - \mu(x) \right) \tau_z + V_z \sigma_z + i \alpha \partial_x \sigma_y \tau_z + \Delta \tau_x
\ee
where $V_z$ is the Zeeman energy, $\Delta$ is the induced superconducting gap, $\alpha$ gives the strength of the 
spin-orbit coupling, and we have suppressed factors of $\hbar^2/(2m)$ in the kinetic term.  
We will take all parameters except the chemical potential to be constant along the length of the wire, and suppose that $\mu(x)$ has a sharp jump (which we will approximate as a step function) at each of the two junctions separating the TSC and NTSC segments of wire.  

In the presence of an interface between TSC and NSC regions of the wire, normalizeable zero energy eigenfunctions of (\ref{HSC}) exist.  The equation describing these zero-mode solutions can be expressed:
\be \label{LittleHam}
\begin{pmatrix} - \partial_x^2  - \mu(x) + V_z& \Delta + \alpha \partial_x \\
- \Delta - \alpha \partial_x &  - \partial_x^2  - \mu(x) - V_z
\end{pmatrix}  u = 0
\ee
where 
\be
\psi^T = ( u_1, u_2, u_2, - u_1 )
\ee

We take the chemical potential  to be
\be
\mu(x) = \begin{cases} \infty & x< x_1 \\
\mu_1 & x_1 <x < x_2 \\
\mu_2 & x_2 < x < x_3 \\
\mu_1 & x_3 < x < x_4 \\
\infty & x> x_4
\end{cases}
\ee 
where $\mu_1$ is chosen so that the wire is in the topological superconducting phase, and $\mu_2$ is chosen such that it is in the normal superconducting phase.  
In each region, the zero mode solutions take the form
\be 
u = u_L + u_R
\ee
with
\be \label{Usum}
u_L = \sum_n  a_{n,L} e^{ - k_n (x - x_L) } u^{(0)}_{L} \ , \  \ \ u_R = \sum_n  a_{n,R}  e^{  k_n (x - x_R) } u^{(0)}_{R}
\ee
(For example, in the left-hand wire segment $x_L = x_1, x_R = x_2$; in the middle (non-topological) segment, $x_L = x_2, x_R = x_3$, and son on).  
Here $k_n$ satisfy the conditions
\be \label{kEq}
(k_n^2 + \mu_i)^2 - V_z ^2 + (k_n \alpha - \Delta)^2 =0 \ , \ \ \ \text{Re}(k_n ) >0
\ee
such that $u$ is a superposition of the bound state wave functions at the left and right ends of the wire segment.  
Eq. (\ref{LittleHam}) is satisfied if $u^{(0)}_{ L, R}$ in Eq. (\ref{Usum}) are given by
\be
u^{(0)}_{L} =\begin{pmatrix} k_n \alpha - \Delta \\ -k_n^2 - \mu_i + V_z \\ \end{pmatrix} \ \ \ 
u^{(0)}_{R} =\begin{pmatrix} - k_n \alpha - \Delta \\ -k_n^2 - \mu_i + V_z \\ \end{pmatrix}
\ee
The coefficients $a_n$ must be chosen such that $\psi$ and its first derivative are continuous across the boundaries separating each region.

The detailed calculation is not required here; instead, we will focus on the qualitative difference between the couplings in the topological and non-topological superconducting segments.  The important point is this: Eq. (\ref{kEq}) has four solutions in the complex plane.  However, we will take $a_{n, L}$ in Eq. (\ref{Usum}) to be non-vanishing only when $k_n$ has a positive real part -- which is a good approximation provided that the wire segments are not too short.  In this limit the overlap will be dominated by the terms with the slowest decay along the wire, meaning the smallest real component of $k_n$.  In the topological regime, these values of $k_n$ are virtually always complex, such that the long-distance behavior of the wave-function is oscillatory.  In the normal superconducting regime, however, the dominant $k_n$ is complex only for relatively large values of the spin-orbit coupling, and oscillations are not generic.

As explained in Ref. \onlinecite{DasSarmaSmoking}, the couplings $t_{ij}$ are given by the matrix elements of the Hamiltonian (\ref{HSC}) between the left and right  bound states.  
Qualitatively, when the dominant long-distance behaviour of the bound-state solutions is oscillatory, then there are oscillations in $t_{ij}$ as a function of $\mu$.  In this situation, $t_{23}$ can in principle be tuned to 0 by a relatively small adjustmant in either $\mu$ (estimated to be on the order of $0.5 \, meV$ by Ref. \onlinecite{DasSarmaSmoking}) or the magnetic field.  However, in the non-topologial wire segment such oscillations are not guaranteed to occur, and there are parameter regimes where the voltage required to tune $t_{23}$ to 0 would be significantly higher.


\bibliography{MeasBib}

\end{document}